\documentclass[11pt]{article}
\usepackage{moriond_webber,epsfig}
\def\as{\alpha_{\mbox{\tiny S}}}
\def\ee{e^+e^-}
\def\ycut{y_{ij}}
\def\yini{y_1}
\begin{document}
\hfill Cavendish-HEP-00/05

\hfill hep-ph/0005035
\vspace*{2cm}
\title{COMBINING QCD MATRIX ELEMENTS AND PARTON SHOWERS \footnote{Talk at
XXXV Rencontres de Moriond, Les Arcs, France, March 2000.}}

\author{ B.R. WEBBER }

\address{Theory Division, CERN, 1211 Geneva 23, Switzerland, and\\
Cavendish Laboratory, University of Cambridge, Cambridge CB3 0HE,
U.K.}

\maketitle\abstracts{
A new method for combining QCD matrix elements and parton showers in
Monte Carlo simulations of hadronic final states is outlined.
The aim is to provide at least a leading-order description of
all hard multi-jet configurations together with jet fragmentation
to next-to-leading logarithmic accuracy, while avoiding the most
serious problems of double counting.}

\section{Introduction}

The Monte Carlo simulation of multi-jet final states is
a challenging problem in QCD and important for new physics searches.
Two extreme approaches to this problem can be formulated as follows.
One can use the corresponding matrix elements with bare partons representing
jets. Then one must add a model for conversion of the partons into hadrons;
any realistic model will include parton showering, and hence extra jet
production and potential double counting.  Alternatively, one can
use the parton model to generate the simplest relevant final state
(e.g.\ $\ee\to q\bar q$) and produce addition jets by parton showering.
However, this gives a poor simulation of configurations with several widely
separated jets.

For earlier work on combining these approaches see
refs.~\cite{Seymour:1995df,Andre:1998vh,Corcella:1998rs,Friberg:1999fh,Collins:2000qd}. Here I outline a method~\cite{Catani:2000xx}
in which the matrix element and parton shower
domains are separated at some value $\yini$ of the $k_T$ (Durham)
jet resolution~\cite{Dokshitzer:1991hj}
$$
y_{ij}\equiv 2\min\{E_i^2,E_j^2\}(1-\cos\theta_{ij})/Q^2\;.
$$

The proposed method has the following features:
At $\ycut>\yini$ multi-jet cross
sections and distributions are given by matrix elements
modified by Sudakov form factors.
At $\ycut<\yini$ they are given by parton showers
subjected to a `veto' procedure, which cancels the
$\yini$ dependence of the  modified matrix elements 
to next-to-leading logarithmic (NLL) accuracy.

Note that the procedure does not aim at a complete description
of any configuration
to next-to-leading order (NLO) in $\as$, although this might
be possible after subtracting NLO terms from the Sudakov form
factors (see ref.~\cite{Collins:2000qd}).
The main objective is to describe all hard multi-jet configurations
to leading order, i.e.\ ${\cal O}(\as^{n-2})$ for
$n$ jets, together with jet fragmentation to NLL accuracy,
while avoiding major problems of double
counting and/or missed phase-space regions.

\section{Modified Matrix Elements}
The exclusive $\ee$ $n$-jet fractions at c.m.\ energy $Q$ and $k_T$-resolution
$\yini=Q_1^2/Q^2$ are given to NLL accuracy by~\cite{Catani:1991hj}
\begin{eqnarray*}
R_2(Q_1,Q) &=& \left[\Delta_q(Q_1,Q)\right]^2\\
\label{eq_R3}
R_3(Q_1,Q) &=& 2\left[\Delta_q(Q_1,Q)\right]^2
\int_{Q_1}^Q dq\,\Gamma_q(q,Q)\Delta_g(Q_1,q)\\
\label{eq_R4}
R_4(Q_1,Q) &=& 2\left[\Delta_q(Q_1,Q)\right]^2 \Biggl\{
\int_{Q_1}^Q dq\,\Gamma_q(q,Q)\Delta_g(Q_1,q)
   \int_{Q_1}^Q dq'\,\Gamma_q(q',Q)\Delta_g(Q_1,q')\\
&+&\int_{Q_1}^Q dq\,\Gamma_q(q,Q)\Delta_g(Q_1,q)
   \int_{Q_1}^q dq'\,\Gamma_g(q',q)\Delta_g(Q_1,q')\\
&+&\int_{Q_1}^Q dq\,\Gamma_q(q,Q)\Delta_g(Q_1,q)
   \int_{Q_1}^q dq'\,\Gamma_f(q')\Delta_f(Q_1,q')\Biggr\}
\end{eqnarray*}
etc., where $\Gamma_{q,g,f}$ are $q\to qg$, $g\to gg$ and $g\to q\bar q$
branching probabilities
\begin{eqnarray*}\label{eq_Gq}
\Gamma_q(q,Q) &=& \frac{2C_F}{\pi}\frac{\as(q)}{q}
\left(\ln\frac Q q -\frac 3 4\right) \\
\label{eq_Gg}
\Gamma_g(q,Q) &=& \frac{2C_A}{\pi}\frac{\as(q)}{q}
\left(\ln\frac Q q -\frac{11}{12}\right) \\
\label{eq_Gf}
\Gamma_f(q) &=& \frac{N_f}{3\pi}\frac{\as(q)}{q}
\end{eqnarray*}
and $\Delta_{q,g}$ are the quark and gluon Sudakov form factors
\begin{eqnarray*}
\Delta_q(Q_1,Q) &=& \exp\left(-\int_{Q_1}^Q dq\,\Gamma_q(q,Q)\right) \\
\Delta_g(Q_1,Q) &=& \exp\left(-\int_{Q_1}^Q dq\,
\left[\Gamma_g(q,Q)+\Gamma_f(q)\right]\right)
\end{eqnarray*}
with
$$
\Delta_f(Q_1,Q) = \left[\Delta_q(Q_1,Q)\right]^2/\Delta_g(Q_1,Q)\;.
$$

The Sudakov form factor $\Delta_i(Q_1,Q)$ represents the probability for
a parton of type $i$ to evolve from scale $Q$ to scale $Q_1$ without
any branching (resolvable at scale $Q_1$). Thus $R_2$ is the
probability that the produced quark and antiquark both evolve from
$Q$ to $Q_1$ without branching. More generally, the probability to evolve
from $Q$ to $q\ge Q_1$ without branching (resolvable at scale $Q_1$) is
$\Delta_i(Q_1,Q)/\Delta_i(Q_1,q)$.

In $R_3$, the quark $q$ (or antiquark $\bar q$) evolves from
$Q$ to $Q_1$ without branching, while the
$\bar q$ (or $q$) evolves from $Q$ to $q$, branches,
and the resulting partons evolve from $q$ to $Q_1$ without branching.
\begin{center}\epsfig{file=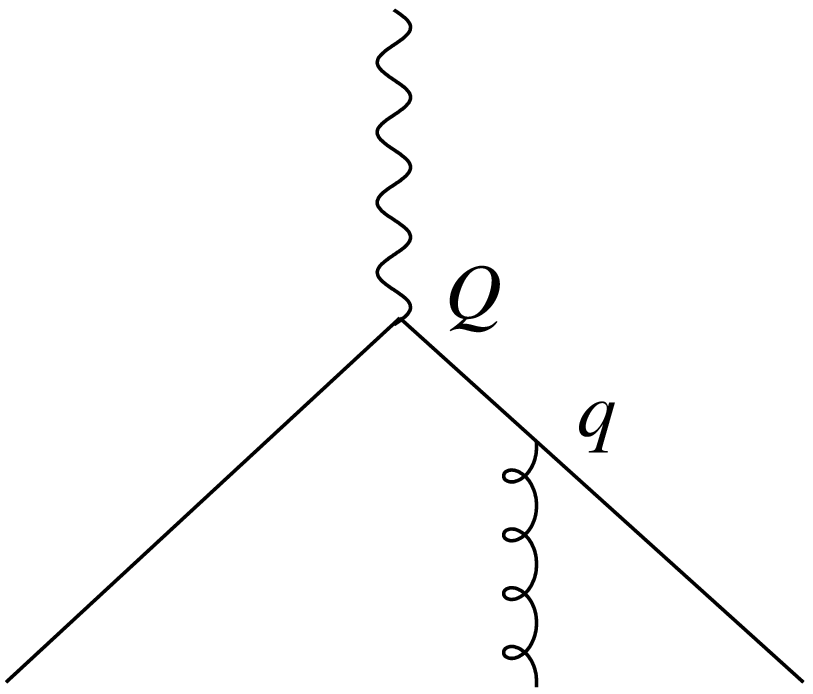,width=4cm}\end{center}
Thus the overall NLL probability is
$$
2\Delta_q(Q_1,Q)\frac{\Delta_q(Q_1,Q)}{\Delta_q(Q_1,q)}
\Gamma_q(q,Q)\Delta_q(Q_1,q)\Delta_g(Q_1,q)\;,
$$
which gives $R_3(Q_1,Q)$ after integration over $Q_1<q<Q$.

We can improve the description of 3-jet distributions throughout
the region $y_{qg},y_{\bar q g}>\yini$ by using the
{\em full tree-level matrix element squared} ${\cal M}_{q\bar qg}$
in place of the NLL branching probability $\Gamma_q(q,Q)$.
We first generate $q\bar q g$ momentum configurations
according to  ${\cal M}_{q\bar qg}$, with $k_T$-resolution cutoff
$\ycut>\yini=Q_1^2/Q^2$,
then weight each configuration with an extra factor of
$[\Delta_q(Q_1,Q)]^2\Delta_g(Q_1,q)$
where $q^2=\min\{y_{qg},y_{\bar q g}\}Q^2$.
For consistency we also use $\as(q)$ in ${\cal M}_{q\bar qg}$.

For four or more jets,
there are several branching configurations with different
colour factors.  For example there is a contribution from
$q\to q g$ branching at scale $q$ followed by $g\to g g$
at scale $q'$:
\begin{center}\epsfig{file=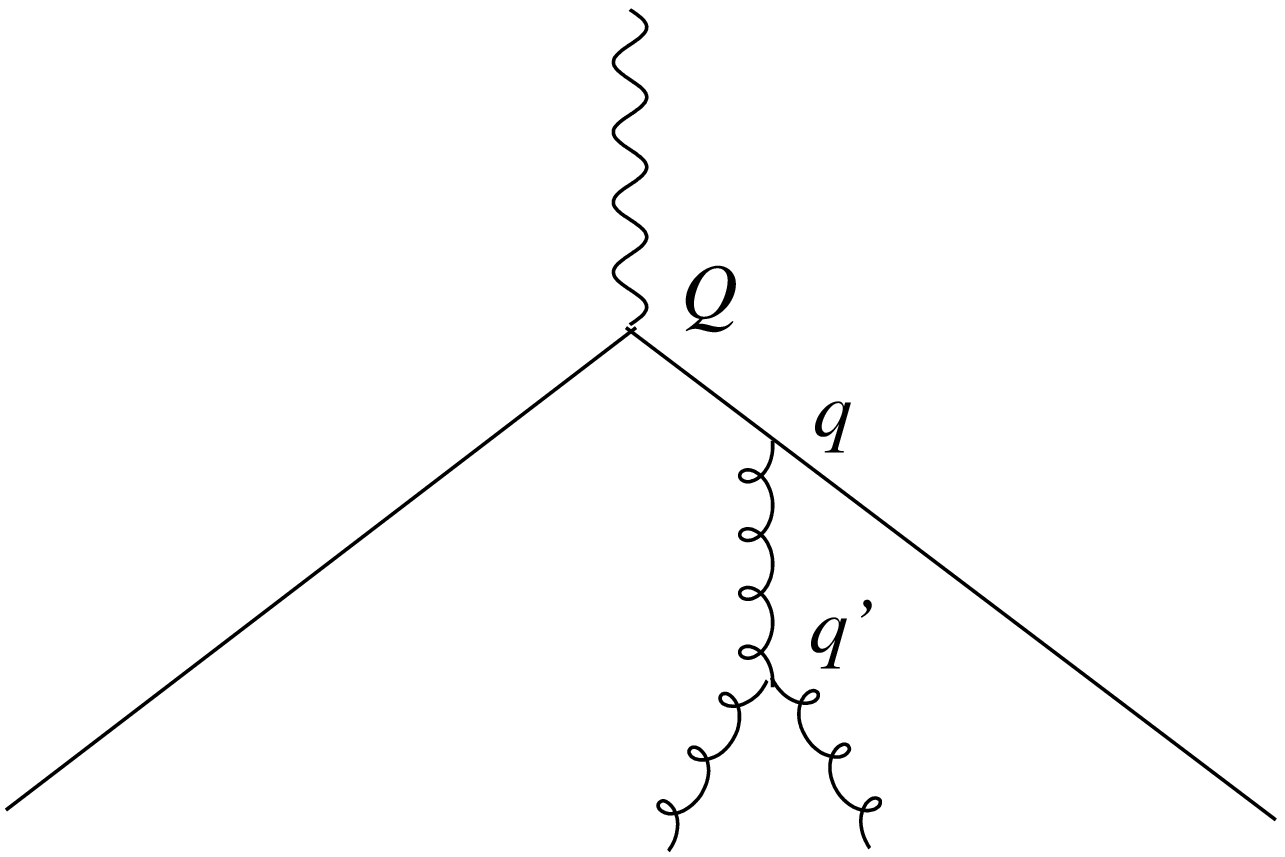,width=5cm}\end{center}
The probability of this is
\begin{eqnarray*}
&&\Delta_q(Q_1,Q)\frac{\Delta_q(Q_1,Q)}{\Delta_q(Q_1,q)}\Gamma_q(q,Q)
\Delta_q(Q_1,q)\frac{\Delta_g(Q_1,q)}{\Delta_g(Q_1,q')}
\Gamma_g(q',q)[\Delta_g(Q_1,q')]^2
\end{eqnarray*}
which contributes to the term with colour factor $C_FC_A$.
The product $\Gamma_q(q,Q)\Gamma_g(q',q)$ is an approximation
to the full matrix element squared ${\cal M}_{q\bar qgg}$
in  the kinematic region where $y_{gg}$ is smallest interparton
separation. Thus it is legitimate in NLLA to replace it by
${\cal M}_{q\bar qgg}$ in that region.  The remaining factor
of  $[\Delta_q(Q_1,Q)]^2\Delta_g(Q_1,q)\Delta_g(Q_1,q')$ is
the extra Sudakov weight to be applied.

In general, the proposed procedure for generating $n$-parton
configurations is thus as follows:
\begin{itemize}
\item
First distribute the parton momenta according to the relevant
$n$-parton matrix element squared ${\cal M}_n$, using a fixed
value $\as(Q_1)$ for the strong coupling.
\item
Use the $k_T$-clustering algorithm to determine the
resolution values $y_2=1>y_3>\ldots,>y_n>\yini$ at which $2,3,\ldots,n$
jets are resolved. These give the {\em nodal values} of $q_j=Q\sqrt{y_j}$ 
for a tree diagram that specifies the $k_T$-clustering sequence for that
configuration.
\item Apply a coupling-constant weight factor of
$\as(q_3)\as(q_4)\cdots\as(q_n)/[\as(Q_1)]^{n-2}<1$.
\item
For each internal line of type $i$ from a node at scale $q_j$ to $q_k<q_j$,
apply a Sudakov weight factor $\Delta_i(Q_1,q_j)/\Delta_i(Q_1,q_k)<1$.
For an external line from a node at scale $q_j$, the weight factor
is $\Delta_i(Q_1,q_j)$.
\end{itemize}
Since the weight factors are all less than unity, unweighted events can
be generated by rejecting those for which the product of weights is
less than a random number.

As an example, the following clustering sequence for $\ee\to q\bar q ggg$
\begin{center}\epsfig{file=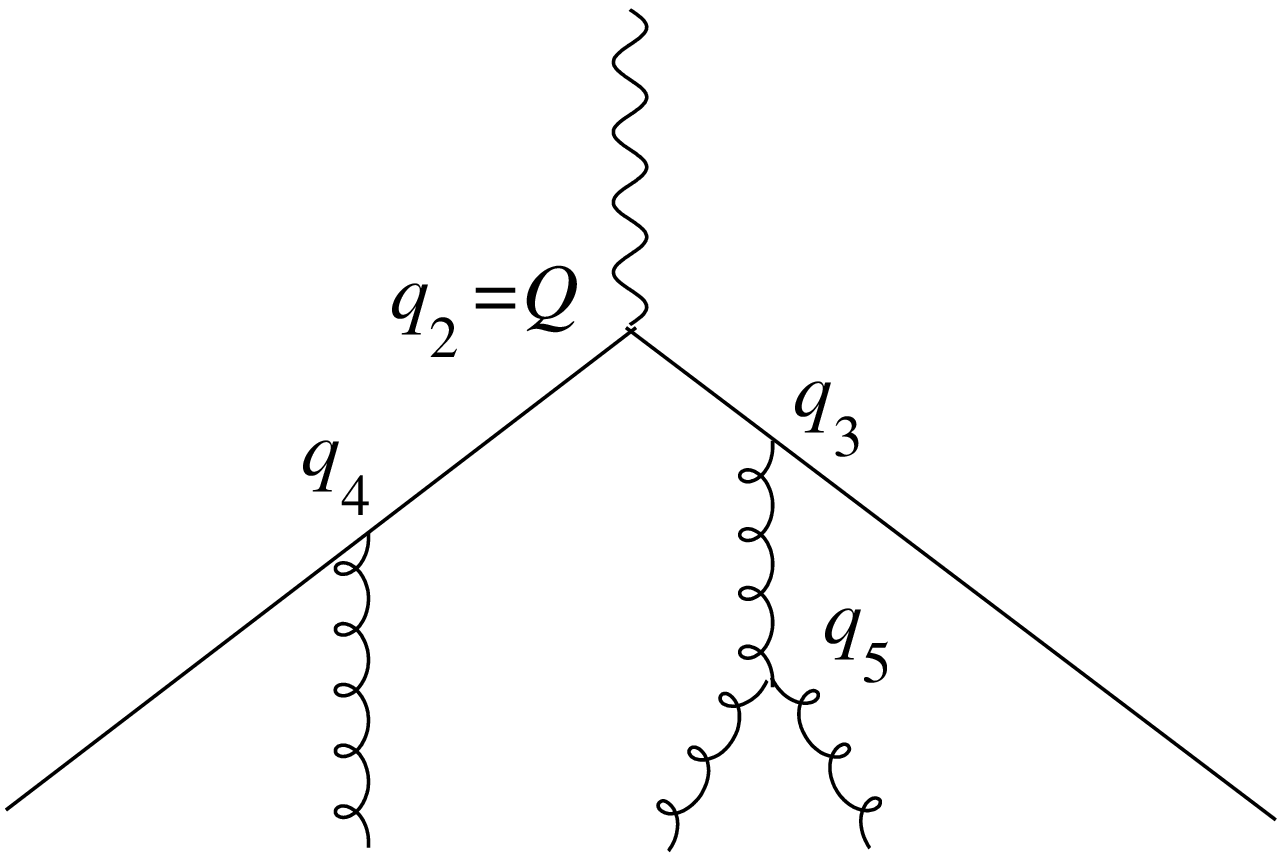,width=5cm}\end{center}
has Sudakov weight
\begin{eqnarray*}
\frac{\Delta_q(Q_1,Q)}{\Delta_q(Q_1,q_3)}
\frac{\Delta_q(Q_1,Q)}{\Delta_q(Q_1,q_4)}
\frac{\Delta_g(Q_1,q_3)}{\Delta_g(Q_1,q_5)}
&\times&\Delta_q(Q_1,q_3)\Delta_q(Q_1,q_4)
\Delta_g(Q_1,q_4)[\Delta_g(Q_1,q_5)]^2 \\
&=&[\Delta_q(Q_1,Q)]^2\Delta_g(Q_1,q_3)\Delta_g(Q_1,q_4)
\Delta_g(Q_1,q_5)
\end{eqnarray*}
Note that the weight factor is actually independent of the structure
of the clustering tree and is the same as that for the Abelian
(QED-like) graph with the same nodal scale values $\{q_j\}$.
The clustering of partons will sometimes be `wrong' but this should
not affect LL and NLL terms. Other clustering procedures can be
envisaged~\cite{Friberg:1999fh} which should be equivalent in the
dominant regions.

\section{Vetoed Parton Showers}
Having generated multijet distributions above the resolution value
$\yini$ according to matrix elements modified by form factors,
it remains to generate distributions at lower values of $\ycut$
by means of parton showers. This should be done in such a way that
the dominant (LL and NLL) dependence on the arbitrary parameter $\yini$
cancels. Any residual dependence on $\yini$ could be useful for tuning less
singular terms to obtain optimal agreement with data.

Note that  $\yini$ must set an upper limit on interparton
separations $y_{ij}$ generated in the showers. Otherwise the
exclusive jet rates at resolution $\yini$ could be changed
by showering.
At first sight, this might suggest that we should
evolve the showers from the scale $Q_1=Q\sqrt{\yini}$
instead of $Q$. However, this would not lead to cancellation of
dependence on $\log\yini$.

Consider, for example, the 2-jet rate at resolution $y_0=Q_0^2/Q^2<\yini$.
If we start from $R_2$ at scale $Q_1$ and then evolve from $Q_1$ to $Q_0$,
we obtain a 2-jet rate of $\left[\Delta_q(Q_1,Q)\Delta_q(Q_0,Q_1)\right]^2$
instead of the correct result
$R_2(Q_0,Q) = \left[\Delta_q(Q_0,Q)\right]^2$.
This is because, although $y_{ij}$ values in the showers
are limited by  $\yini$, the {\em angular regions} in which
they evolve should still correspond to scale $Q$ rather than $Q_1$.
Consequently we should allow the showers to evolve from scale $Q$
but {\em veto} any branching with scale $q>Q_1$ -- i.e., the selected
parton branching is forbidden but that parton has its scale reset
to $q$ for subsequent branching.

The 2-jet rate at any scale $Q_0<Q_1$ is now
given by the sum of probabilities of $0,1,2,\ldots$ vetoed branchings
(represented by crosses) and no actual resolved branchings:
\begin{center}\epsfig{file=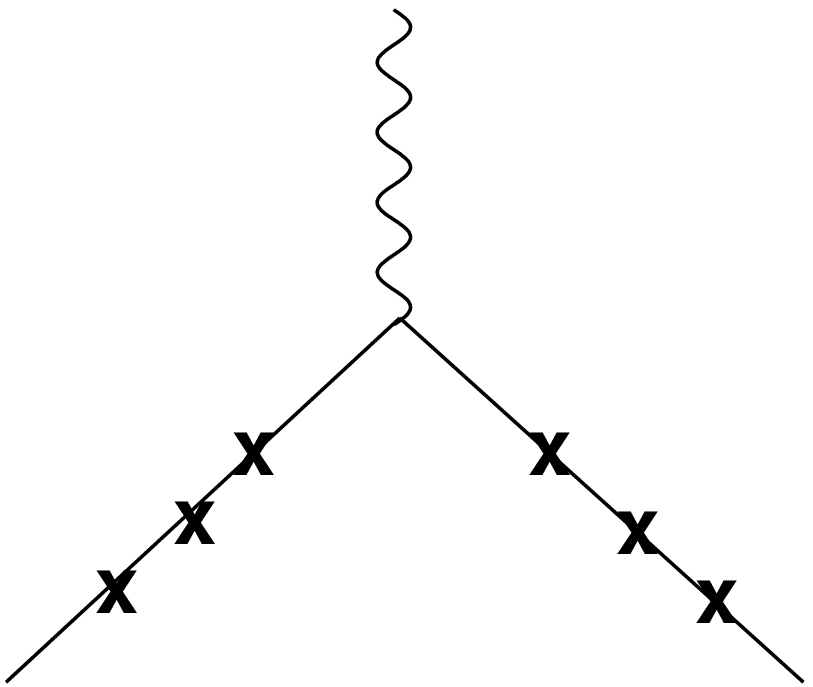,width=4cm}\end{center}
The sum of these probabilities for the quark line is
\begin{eqnarray*}
&&\Delta_q(Q_1,Q)\Delta_q(Q_0,Q)\Biggl\{1+\int_{Q_1}^Q dq\,\Gamma_q(q,Q)
+ \int_{Q_1}^Q dq\,\Gamma_q(q,Q)\int_{Q_1}^q dq'\,\Gamma_q(q',Q)
+\cdots\Biggr\}
\end{eqnarray*}
The series sums to $1/\Delta_q(Q_1,Q)$, cancelling the $\yini$ dependence
and giving $\Delta_q(Q_0,Q)$. Similarly for the antiquark line.

For the 3-jet rate at scale $Q_0<Q_1$ there are two possibilities:
either the event is a 2-jet at scale $Q_1$ and then has one branching
resolved at scale $Q_0$, or it is a 3-jet at scale $Q_1$ and remains
so at scale $Q_0$. The probability of the first case is
\begin{center}\epsfig{file=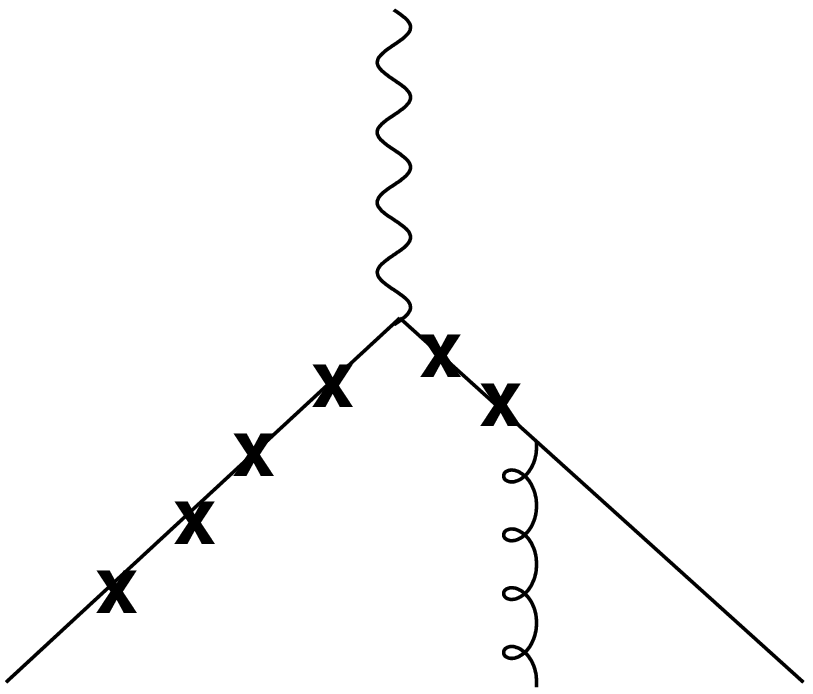,width=4cm}\end{center}
$$
2[\Delta_q(Q_1,Q)]^2\left[\frac{\Delta_q(Q_0,Q)}{\Delta_q(Q_1,Q)}\right]^2
\int_{Q_0}^{Q_1} dq\,\Gamma_q(q,Q)\Delta_g(Q_0,q)
$$
while that of the second case is
\begin{center}\epsfig{file=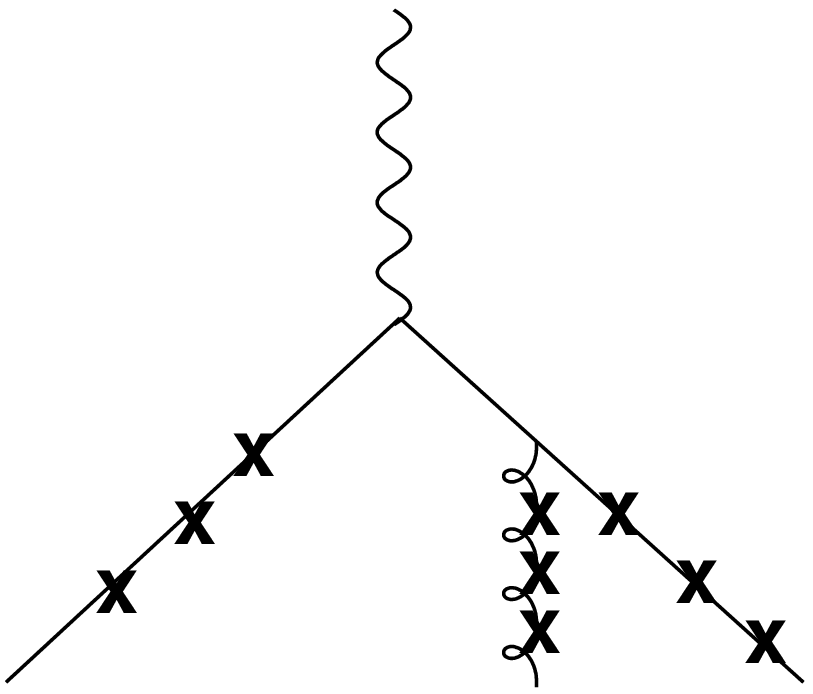,width=4cm}\end{center}
\begin{eqnarray*}
&&2[\Delta_q(Q_1,Q)]^2\left[\frac{\Delta_q(Q_0,Q)}{\Delta_q(Q_1,Q)}\right]^2
\int_{Q_1}^Q dq\,\Gamma_q(q,Q)\Delta_g(Q_1,q)
\frac{\Delta_g(Q_0,q)}{\Delta_g(Q_1,q)}\;.
\end{eqnarray*}
The sum is indeed $\yini$-independent and equal to $R_3(Q_0,Q)$.
Similarly for higher jet multiplicities.

Notice that evolution after a branching at scale $q>Q_1$
starts at scale $q$ rather than $Q$ or $Q_1$. In general, vetoed showers
should evolve in the {\em phase space for angular-ordered branching}
of each parton~\cite{Marchesini:1988cf}.
This depends on the colour structure of the matrix element.
As illustrated below,
the angular region for parton $i$ is a cone bounded by the direction of
parton $j$ (and vice-versa), where $i$ and $j$ are colour-connected.
\begin{center}\epsfig{file=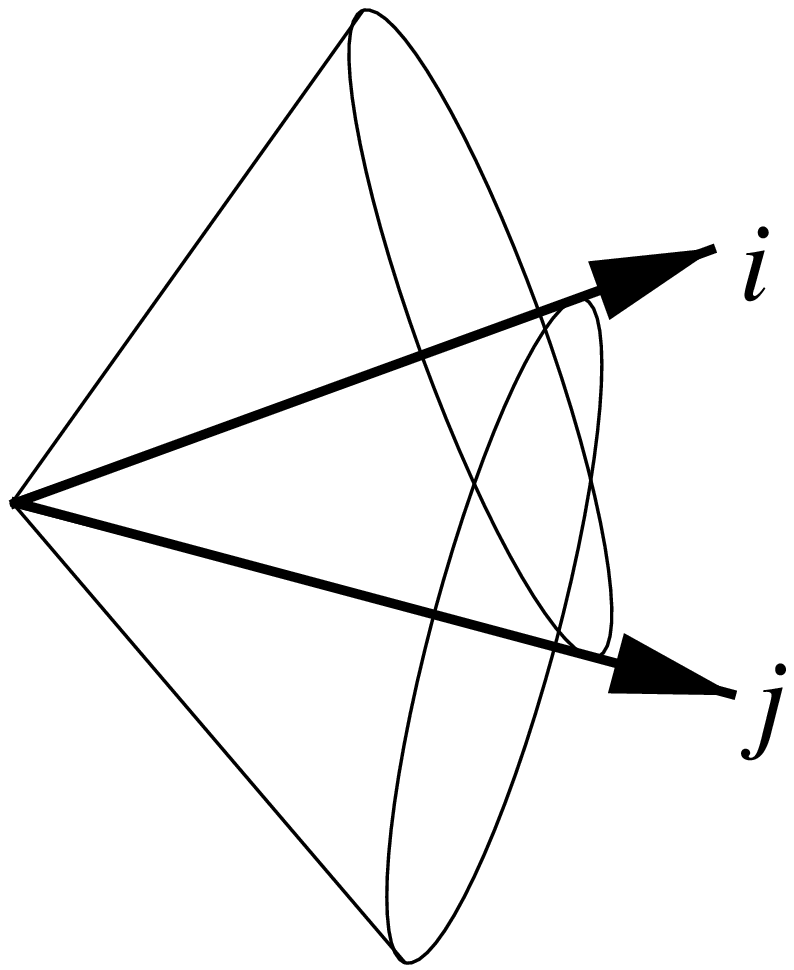,width=4cm}\end{center}

If the colour structure is not unique, colour connections must be
selected according to their relative contributions to the matrix
element squared, which are well-defined in the limit that the number
of colours $N_c$ is large. Corrections to the large-$N_c$ limit are normally
of relative order $1/N_c^2$, so this approximation is adequate to $\sim 10\%$.
For high parton multiplicity, when the colour structure is not known
even at large $N_c$, it may be possible to use the clustering scheme
discussed above as a first approximation in assigning colour connections. 

\section{Comments/Conclusions}
\begin{itemize}
\item Modified matrix elements plus vetoed parton showers, interfaced
at some value $\yini$ of the $k_T$-resolution parameter, should provide
a convenient way to describe simultaneously the hard multi-jet and
jet fragmentation regions.
\item The matrix element modifications are coupling-constant
and Sudakov weights computed directly from the $k_T$-clustering sequence.
\item Dependence on $\yini$ is cancelled to NLLA by
vetoing $y_{ij}>\yini$ in the parton showers.
\item  This prescription avoids double-counting problems
and missed phase-space regions.
\item In principle one needs the matrix elements
${\cal M}_n$ for $\ycut>\yini$ at {\em all values of $n$}.
In practice, if we have $n\le N$, then $\yini$ must be chosen
large enough for $R_{n>N}(Q_1,Q)$ to be negligible.
\item This approach is being implemented (with $N=5$)
in the $\ee$ event generator {\small APACIC++}~\cite{Krauss:1999fc}.
\item It should be possible to extend it to lepton-hadron and  
hadron-hadron collisions.
\item Extension to NLO along the lines of ref.~\cite{Collins:2000qd}
may also be possible.
\end{itemize}

\section*{Acknowledgments}
It is a pleasure to acknowledge my collaborators in this project,
S.~Catani, F.~Krauss and R.~ Kuhn.

This work was supported in part by the UK Particle Physics and
Astronomy Research Council and by the EU Fourth Framework Programme
`Training and Mobility of Researchers', Network `Quantum Chromodynamics
and the Deep Structure of Elementary Particles',
contract FMRX-CT98-0194 (DG 12 - MIHT).

\end{document}